\documentclass[runningheads]{llncs}
\usepackage{graphicx}

\usepackage{booktabs}       
\usepackage{amsfonts}       
\usepackage{nicefrac}       
\usepackage{microtype}      
\usepackage[table,xcdraw]{xcolor}
\usepackage{fontawesome}
\usepackage{multicol}
\usepackage{amsmath}
\usepackage{subfigure}
\usepackage{float}
\usepackage{listings}
\usepackage{arydshln}
\usepackage{circuitikz}

\usepackage{tikz}
\usetikzlibrary{intersections}
\usepackage{xcolor}
\usepackage{parcolumns}
\usepackage{listings}
\usepackage{caption}

\newcommand{\lstbg}[3][0pt]{{\fboxsep#1\colorbox{#2}{\strut #3}}}
\lstdefinelanguage{diff}{
  basicstyle=\ttfamily\footnotesize,
  morecomment=[f][\lstbg{red!20}]-,
  morecomment=[f][\lstbg{green!20}]+,
  morecomment=[f][\textit]{@@},
}

\begin{document}
\title{On Plagiarism and Software Plagiarism}
\author{Rares Folea\inst{1,2}\orcidID{0000-0002-4936-9082} \and Emil Slusanschi\inst{1}\orcidID{0000-0003-0222-1002}}
\authorrunning{R. Folea and E. Slusanschi}

\institute{Department of Computer Science and Engineering, Faculty for Automatic Control and Computers,\linebreak
National University of Science and Technology Politehnica Bucharest, Romania \and Doctoral School of Engineering and Applications of Lasers and Accelerators (S.D.I.A.L.A.)}
\maketitle 
\begin{abstract}
This paper explores the complexities of automatic detection of software similarities, in relation of the unique challenges of digital artifacts and introduces Project Martial, an open-source software solution for detecting code similarity. This research enumerates some of the existing approaches to counter software plagiarism, by examining both the academia and legal landscape, including notable lawsuits and court rulings that have shaped the understanding of software copyrights infringements for commercial use applications. Furthermore, we categorize the classes of detection challenges, based on the available artifacts, and we provide a survey on the previously studied techniques in the literature, including solutions based on fingerprinting, software birthmarks or code embeddings and exemplify how a subset of them can be applied in the context of Project Martial.

\end{abstract}

\section{Introduction}

In its most general definition, \textit{plagiarism}\footnote{The word originates from the Latin \textit{"plagiarius"}, meaning torturer, oppressor or kidnapper.} is the act of presenting another person's work as original, without proper attribution. While there were many cases of \textit{plagiarism} in history, the first documented case of using the word \textit{plagiarius}~\cite{seo2009plagiarism} was during the life of the roman poet Martial, when exact copies of his poems and epigrams started to appear presented as personal work authored by some other, obscure writers. He is considered to be the first person to claim authorship rights, in an ancient world where intellectual property and copyright laws not only were not enforced, but not even considered.

\textbf{Software plagiarism} is just one of the many branches of plagiarism, but due to the nature of software development, which is an almost entirely digital, fast-paced environment, where duplication activities can be performed really fast, the number of acts related to software plagiarism is raising swiftly. The main motivation behind plagiarizing software is to avoid the effort required to develop a novel, original solution from scratch.

Software plagiarism is a widespread issue affecting both academic world – where students might seek better grades – and commercial space, where intellectual property lawsuits can raise claims for up to several billions of dollars. Despite the vastly different stakes, the core act of plagiarism remains the same: illegitimately claiming authorship of another person's work. 

Famous cases of plagiarism attract significant public attention, and software plagiarism is no exception. Remarkably, these debates and allegations might extend beyond the field of software engineering. Some high-profile allegations have been dramatized in films like \textit{"Browser Wars"} and \textit{"The Billion Dollar Code"}. In contrast, cases of alleged plagiarism in introductory computer science courses may topics of discussion among students, sparking conversations about ethics and decisions of the university committees.

In \textbf{Section~\ref{section-comparison}} of this paper, we explore the unique aspects of software plagiarism compared to other fields. \textbf{Section~\ref{section-academia}} examines plagiarism's impact within academia, while \textbf{Section~\ref{section-legal}} addresses legal implications associated with plagiarism. \textbf{Section~\ref{section-classes}} aims to provide a categorization of possible classes of solutions that may arise to detect plagiarism, based on the availability of source code or binary code, and in \textbf{Section~\ref{section-techniques}}, we present certain techniques that have been used for automatic detection. Finally, in \textbf{Section~\ref{section-future-work}} we discuss promising areas for future research and we introduce Project Martial, an open-source software solution that helps detecting code similarities.

\section{Comparison with other kinds of plagiarism}
\label{section-comparison}

How is detecting software plagiarism different than other kinds of plagiarism is a good, yet difficult to answer question. It is hard to believe that a trivial set of rules can cover all the possible situation, because of the nature of software development dependencies. As in the  examples mentioned later, where a ruling of the US Supreme Court acknowledged that cases where duplicating the declarative part of the code does without copying the implementation part is not to be considered plagiarism. As a direct consequence, the ruling stated that APIs are not subject to copyrights infringements. This gives a unique characteristic on how plagiarism should be seen in the software industry as opposed to other industries, such as music or literature, where there is no clear correspondence between the declarative and implementing part of the work.

In general, \textit{there is no consensus}~\cite{chuda2011issue} \textit{around the difficulty of the problem}, with many multiple arguments pro- and against- the ease of detecting plagiarism in software as opposed to identifying it in other work.

On one hand, there are arguments supporting that similarity between two sources written in natural language text rather than software is much easier to be reasoned for humans. That is because, in the case of human language, references and quotes in the corpus may lead to texts of $10\%$ similarity that are likely not be considered plagiarised, while for software development, analogous working methods may be hard to be labeled, because the distinction between "\textit{same theoretical knowledge}" and "\textit{plain copy}" is more difficult to be established, especially in situations where there are a very few ways of getting things right~\cite{rosales2008detection}. A very famous duplication of code without raising the question of plagiarism was presented by Ken Thompson during 1984's Turing Award Lecture~\cite{thompson1984reflections}, when citing his collaboration work with Dennis Ritchie, he mentioned that towards their long collaboration, only one occasion of miscoordination of work has happened. Thompson outlines that he discovered that both have \textit{"written the same 20-line assembly language program"}. He concludes: \textit{"I compared the sources and was astounded to find that they matched character-for-character"}.

On the other hand, arguments that that software plagiarism may be easier identifiable are centered around the language complexity between the two, with~\cite{lancaster2005classifications} reasoning that a text in a human language contains an effectively unlimited number of possible words that can be used, with the both intra- or extra-corporal plagiarism methods, while code plagiarism is constrained to a well defined set of keywords, provided by the minimal vocabulary of a given programming language. 
 
To add some complexity to this problem, one additional constraint arise when there is a scenario with no access to the source code of the suspicious programs, and the only available resource is the executable binary. Because the assembly instructions are far more obfuscated for a human-mind structured analysis, trying to evaluate the originality of an executable is a substantially more challenging problem than the study of source code~\cite{zhang2014program}. 

Nevertheless, not providing the source code is not the only possible way to obfuscate the analysis. This can also be achieved, by applying iterative changes to the original code, that do not modify the behaviour of the program, but makes it look different. There are numerous obfuscation techniques that have been developed, that can be mainly classified into both control- and data-oriented; starting from naive techniques such as symbols renaming or boolean splitters (refer to the example in Listing), and advancing to more involved changes, such as adding noise instructions, reordering independent instructions, loops unrolling (Listing~\ref{loop-unrolling}), changes in the control flow conditional branches, branch inversion (Listing~\ref{control-flow-conditional-branches}) or aggregating multiple instruction (Listing~\ref{aggregating-multiple-instruction}). The examples, provided in the syntax of the Go programming language, are just some simple academic samples that can alter~\footnote{It is worth mentioning that some modifications (such as loop-unrolling or vectorisation) have multiple functionality. These techniques can be used not only as methods for obfuscation, but rather they can serve as optimization techniques, in order to make the code more efficient, rather than trying to make it look different. However, in the scope of the paper only modifications with the goal of producing code obfuscations are being studied.} the original code in a way that the resulting code look substantially different.

To add, studies~\cite{madou2006effectiveness,wu2010mimimorphism} have investigated solutions for adding a set of obfuscations who are mainly aimed at protecting the intellectual property, encapsulated by the software, by making it extremely hard for a person that only has access to the obfuscated outcome to decode the original logic. Such perplexity techniques can be applied both at source code level, as well as at binary level. Just as the presented techniques presented above, because the outcome of these procedures is to alter the outlook of the software without altering the behaviour, these can also can be used to abuse software plagiarism detection tools into miss detecting similarity~\cite{tian2015software}.

Special considerations about software plagiarism\footnote{Large language models capabilities perform well in more assignments, such as free form natural language essays, not just on altering software programs, which introduce an additional challenge in other areas of plagiarism too.} have to be raised in the light of the recent evolution of LLM-based solutions, that proved to be efficient at manipulating large code snippets\footnote{Fortunately, there is also a bright side when it comes to new LLM-based tools. There is evidence that these tools help~\cite{dell2023navigating} improving the productivity by assisting developers (and not only). OpenAI's GPT-4~\cite{openai2023gpt4}, Google's Gemini~\cite{GeminiGo65-online} and other tools can nowadays generate hundred of lines of code from a simple prompt.}. Large pretrained language models have already been used and were proved to be efficient in altering the original source code, such that it has remained undetected~\cite{biderman2022fooling}.

This is detailed in~\cite{khalil2023will} which concludes that LLMs~\textit{"have a great potential to generate sophisticated text outputs without being well caught by the plagiarism check software"}. Finally, preliminary results using this new technology seems to suggest that pre-LLM tools have only \textit{"scratched the surface of the possibilities compared to what large neural language models can achieve in producing convincing high-quality paraphrases"}~\cite{wahle2022large} and acknowledge that even humans ability to detect machine-based obfuscations \textit{"appear to decrease with increasing model size} as \textit{they can change sentence structure and word order instead of single word replacements"}.

\section{Academia}
\label{section-academia}

In the academic world, plagiarism, seen as \textit{"an evasion of learning"}\cite{warn2006plagiarism}, because it \textit{"enables students to gain credit for significant portions of assessment without having developed any capacity for understanding of critical evaluation of the material presented for assessment."}, is not a seldom event. As the number of computer science students is still increasing each year, new challenges arise. Data USA states that the total number of degrees awarded in $2020$ exceeded fifty-thousand graduates~\cite{datausa}, with a 10\% growth rate year-to-year, and The British Computer Society noted that~\cite{bcs} the applications to study computer science in United Kingdom for $2022$ increased from $140$ thousands the previous year to almost $160$ thousands this year, with a trend that shows that interest in the computer science related subject kept growing.

Given the scale, there is definitely a need for improving the automatic tools that allow for the identification of possible cases of software plagiarism. As~\cite{chuda2011issue} notes, "\textit{there are too many students and too few staff members [...]. Automating plagiarism detection would help very much}".  

To begin with, various works~\cite{foltynek2020testing} acknowledges the difficulties for identifying plagiarism in natural language for essays and reports with existing tools, yet software plagiarism has it's own specificity. A survey on the topic of the academic issue associated with software plagiarism~\cite{chuda2011issue} presents that only 8\% of the staff university did use a automatic software-based approach to try and detect plagiarism, while the vast majority of 80\% of the staff still relies on instinct and/or personal experience to tackle this problem. It has been pointed out that the complexity in understanding the nature of plagiarism could result in having solutions \textit{"to reducing plagiarism that may rest more on prevention rather than detection"}~\cite{warn2006plagiarism}.

The most used system for automatic software plagiarism detection is Moss\footnote{Moss stands for Measure Of Software Similarity and is an automatic system for determining code similarity. \url{https://theory.stanford.edu/\%7Eaiken/moss/}}. The maintainers present it as \textit{"a way of highlighting the components of programs that are worth a more thorough inspection, that saves instructors and teaching staff a lot of time"}~\cite{Plagiari3:online}. It's emphasised that \textit{"after specific sections of the programs have been examined, it shouldn't matter whether the questionable code was initially uncovered by Moss or by a person"} and that \textit{"the argument for plagiarism should stand on its own"}. Multiple researchers subscribed to this ethic idea and as~\cite{wahle2022large} points out, while automatic plagiarism detection can \textit{"point out potential plagiarism cases, a team of experts should make a final decision on such cases"}, because, the same work in~\cite{wahle2022large} warns that \textit{"False-positive cases of wrongly accused researchers could ruin their careers forever} and therefore, \textit{"all cases should be carefully evaluated before any final verdict"}.

It is also worth pointing to research such as Mossad~\cite{10.1145/3428206}, that showed that automatic solutions to generate mutants of software versions with the sole purpose of defeating plagiarism detectors are possible. 

The core difference between academia and industry is the amount of involvement that the actors are motivated to invest, in the process of plagiarising. While in the commercial context there may be a lot more resources and incentives available, if it is necessary to defend against copyright violation, in an educational context, \textit{"the effort expended by students to hide their plagiarism is likely to be much less"}~\cite{joy1999plagiarism}.

\section{Legal Aspects and Lawsuits}
\label{section-legal}

The fast development of the Tech industry led to the development of a relatively new area in Legal, specialised in technology-based litigation. The majority of lawsuits are centered towards determining the eligibility of a given trademark or patent. Well known trademark cases include both lawsuit with hardware-established claims, such as Apple has demanded ownership claim of its iPad trademark in China, but software claims are no exception and, in the corporate world there are many such well known cases and accusations of software plagiarism. 

Software copyright- and plagiarism-allegations are seldom straightforward to decide, and most of them requires a trial to reach a verdict and in some special cases, even involvement of the US Supreme Court. 

\textit{Oracle America Inc v. Hewlett Packard Enterprise Co} case has been settled between the parts after a "\textit{disputed unauthorized software updates for the Solaris operating system owned by Oracle}", with alleging accusations that Hewlett Packard Enterprise "\textit{used its copyrighted software without a license and undercut Oracle's support pricing}"~\cite{hpeo}, with the goal of reducing the cost of new software development. The telecommunication company \textit{Verizon}, according to the lawsuit against it, has allegedly utilized \textit{BusyBox} programming in wireless routers, which were distributed to consumers, but failed to give customers access to the BusyBox source code as required by the GPL~\cite{Opensour85:online} licence. This case was also settled, subject to license adherence.

In \textit{Google LLC v. Oracle America Inc.}, the US Supreme Court considered multiple factors, pointing out that Google’s use of the Java APIs was transformative and the amount of interfaces copied was limited only to the bare minimum which was necessary to allow developers to work in the new mobile development framework proposed by Android, with the already accumulated knowledge, from the Java programming language. In the preparation of the case, Google argued that \textit{"Open interfaces between programs are the building blocks of many of the services and products we use today, as well as of technologies we haven’t yet imagined. An Oracle win would upend the way the technology industry has always approached the important issue of software interfaces. It would, for the first time, grant copyright owners a monopoly power to stymie the creation of new implementations and applications. And it would make it harder and costlier for developers and startups to create more products for people to use"}~\cite{gg}. On the opposite side, Oracle considered that this trial started \textit{for the simple proposition that stealing—no matter the convenience it may offer to the thief—is not acceptable}\cite{orlc} and "\textit{attempts to claim that its theft and clear commercial use of an existing technology already in the market is somehow covered by the fair use doctrine}", with such an argument "\textit{would virtually eliminate copyright altogether because nearly all copying would be $<<$fair$>>$}"\cite{orlc}.


The jury considered that the portion of copied code was "\textit{inherently bound together with uncopyrightable ideas}" with only $0.4\%$ of the entire API being copied by Google, an argument in favor of \textit{fair use}. Nevertheless, the Android platform was proved by time not to be a substitute or a rival for Java SE. Within the jury there was also a dissenting opinion on this trial, arguing that the Court should have addressed the question whether Oracle’s code is copyrightable and if by copying the code, Google gained multiple advantages and erased a large value of Oracle’s partnerships.

\section{Classes}
\label{section-classes}

\begin{figure}[!ht]
\centering
\resizebox{.6\textwidth}{!}{%
\begin{circuitikz}
\tikzstyle{every node}=[font=\LARGE]
\draw [ fill={rgb,255:red,255; green,163; blue,163} ] (10.25,11.25) circle (3.25cm) node {\LARGE SC.D} ;
\draw [ fill={rgb,255:red,168; green,255; blue,170} , line width=0.5pt ] (4.75,11.5) circle (3.5cm) node {\LARGE SC.S} ;
\draw [ fill={rgb,255:red,255; green,196; blue,148} ] (8.75,9) circle (1.5cm) node {\LARGE BIN.D} ;
\draw [ fill={rgb,255:red,226; green,236; blue,177} ] (5.75,9) circle (2cm) node {\LARGE BIN.S} ;
\draw [, dashed] (4.75,11.5) circle (3.5cm);
\draw [, dashed] (10.25,11.25) circle (3.25cm);
\draw [, dashed] (8.75,9) circle (1.5cm);
\draw [, dashed] (5.75,9) circle (2cm);
\end{circuitikz}
}%
\caption{The overlap of techniques, based on the class of problem.}
\label{classification-venn}
\end{figure}
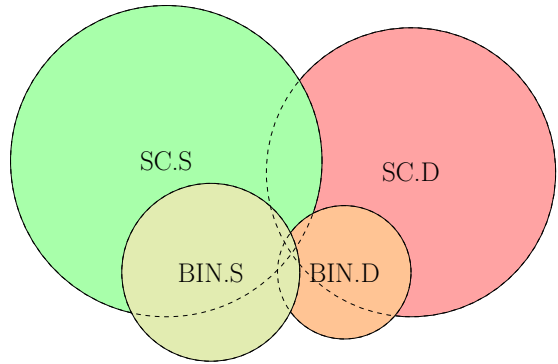

As mentioned in the previous sections, the problem of identifying software plagiarism can require substantially different approaches based on the available resources. We identify two major sets of problems based on that: the class of detection with access to the source code (\textbf{SC}) and the class of detection with access to the binary code only (\textbf{BIN}). Note that while access to \textbf{SC} also implies access to the \textbf{BIN} resources, by simply taking the source code and compiling (for languages whose native implementation allows compiling into the machine language) or packing (for interpreted languages), there may be special cases in which the tool used to translate the code to machines is either unavailable, undisclosed or proprietary. This raises a very niche class of problems in which the exact binary cannot be obtained from the source code. 

Naturally, the semantics of the code can be much easier analysed when access to the source code is available, as the description of the processes that the program needs to follow is more expressive in a high-level programming language than in binary machine-specific instructions~\cite{cesare2012software}. A visual representation of the overlap of techniques that can be used to tackle each challenge is captured in Figure~\ref{classification-venn}.

Therefore, the \textbf{SC} class of problems may result in better software plagiarism detectors, due to the additional access to the source code, which can be tackled in more advanced ways. For example, the approaches can perform analysis on the static representation of the code \textbf{SC.S}, or dynamically \textbf{SC.D}, by monitoring the execution of the program. 

Similarly, we can define \textbf{BIN.S}, as the class of approaches when the machine code is evaluated directly, using a static approach. Symmetrically, \textbf{BIN.D} can be defined, for classes where the machine code is evaluated directly, dynamically, but it almost perfectly overlaps\footnote{The only distinction that we have found was, when in the transformation from the source code to the binary, debug traces and symbols are being enabled, which could be missing otherwise, if the binary is the only resource available.}. with \textbf{SC.D}. 

The problem of automatic software plagiarism detection is far from reaching a generic solution, but among the years of research in literature, some progress towards identifying an automated way of potential cases of plagiarism was made. The traditional approaches are to only proclaim plagiarism when similarities are found\footnote{While most of the solutions focus on \textit{identifying similarities} and only declare plagiarism when a similarity is found, a different approach is taken in the work of~\cite{zhang2014program}, that starts from the idea that \textit{"as long as we can find one dissimilarity, programs are semantically different"}, yet emphasises that \textit{"if we cannot find any, it is likely a plagiarism case"}. This is a much more ambitious approach, but comes with the drawbacks that it lacks explainability which makes the human analysis proposed in the methodology of~\cite{wahle2022large} significantly more difficult.
}.

\section{Techniques in automatic plagiarism detection}

The \textbf{fingerprint-based} approaches can provide automatic findings of program similarities, but in most of the cases, these methods also lack convincing explainability capabilities in order to determine why certain programs are similar. One well establish algorithm is winnowing\cite{schleimer2003winnowing}, a solution for fingerprinting that divides the file into k-grams and has been the core idea for the MOSS software, but there are many other successful usages of fingerprint ideas in other work too, both on \textbf{SC}~\cite{chilowicz2009syntax,narayanan2012source,cesare2012software,chilowicz2009syntax,narayanan2012source,cesare2012software}, as well as on \textbf{BIN}~\cite{cesare2012software}. In the good faith principle, \textit{it is still up to a person to review the portions of the code that the engine has flagged and make a judgement whether that section is indeed plagiarism}~\cite{Plagiari3:online,wahle2022large} or not.

\label{section-techniques}

\begin{table}
\centering
\begin{tabular}{|l|c|c|c|c|}
\hline
\rowcolor[HTML]{EFEFEF} 
Techniques                                                         & SC.S                      & SC.D                      & BIN.S                     & BIN.D                     \\ \hline
\begin{tabular}[c]{@{}l@{}}Code analysis \\ (syntactic, semantic)\end{tabular} & \checkmark &                           &                           &                           \\ \hline
API-based                                                          & \checkmark &                           & \checkmark &                           \\ \hline
Profiling                                                          &                           & \checkmark &                           & \checkmark \\ \hline
Fingerprint                                                        & \checkmark & \checkmark & \checkmark & \checkmark \\ \hline
\begin{tabular}[c]{@{}l@{}}Software\\ birthmarks\end{tabular}      & \checkmark & \checkmark & \checkmark & \checkmark \\ \hline
\begin{tabular}[c]{@{}l@{}}Code\\ Embeddings\end{tabular}          & \checkmark & \checkmark & \checkmark & \checkmark \\ \hline
\begin{tabular}[c]{@{}l@{}}LLM-based\end{tabular}          & \checkmark &  & & \\ \hline
\end{tabular}
\vspace{1em}
\caption{The main techniques used in plagiarism detection, and known classes of problems where they have been applied to detect plagiarism.}
\label{table-plagiarism}
\end{table}
\vspace{-2em}

\textbf{Software birthmark}\footnote{Birthmarks are generally computed as a set of fairly invariable features that are extracted from the analysed algorithm, with the overall goal of uniquely identifying the program. Once birthmarks are extracted, the similarity between two programs is simply computed by a function that takes the two resulting birthmarks and compares them.} approaches are meant to overcome some downsides of software fingerprinting. Like the fingerprint-based algorithms, birthmarks have the advantage that they can be applied on both \textbf{SC}, source code~\cite{myles2005k,chilowicz2009syntax,cesare2012software,tian2015software,myles2004detecting} as well as on \textbf{BIN}, where only the binary code is provided~\cite{cesare2012software,7153572,tian2015software,ullah2021software,lu2007software}. Birthmark techniques can be enhanced with additional mechanisms in complex programs, such as supervising the construction~\cite{tian2020plagiarism} via Siamese networks to capture perturbations caused by the non-deterministic scheduling of machine instruction in the context of multi-threading applications.

\textbf{Code embeddings} are yet another powerful mechanism that can be used in similarity detection in which the program is analysed via its numerical vector representations. They aim to capture the semantic meaning and syntactic structure of code snippets in a condensed format and have potential in placing the code snippets with similar functionality close to each other in the multi-dimensional analysed space. While most approaches aims to tackle \textbf{SC} problems~\cite{chen2019literature}, they can also be applied for \textbf{BIN}~\cite{folea2023complexity} classes. 

For studying \textbf{SC.S} problems, the studies in~\cite{liu2006gplag,chae2013software} are using a detection method running an \textbf{API-labeled dependency} graph analysis.

A encouraging, but yet to be explored territory is the use of Large Language Models as a technique for automatic plagiarism detection in software. Due to the proved good LLMs performance manipulating source code, this is a promising technology in engaging in \textbf{SC.S} problems.

Note that final applications may use a mix of these approaches from different solutions for analysing similarity. As long as they assume that the problem is in \textbf{SC}, all presented techniques can be applied: \textbf{SC.S}, \textbf{SC.D}, \textbf{BIN.S}, \textbf{BIN.D}. If the problem is in \textbf{BIN}, the available techniques can belong to \textbf{BIN.S} and \textbf{BIN.D}. A summary of the main existing approaches has been captured in Table~\ref{table-plagiarism}.

\section{Project Martial: current and future work}
\label{section-future-work}

Part of the authors' research and contributions to the field of similarity detection is focused around finding novel methods that operate on both the static code, but also perform dynamic analysis (for both \textbf{SC} and \textbf{BIN} classes). Project Martial\footnote{\url{https://github.com/raresraf/project-martial}} is a growing, experimental, open-source initiative, that aims to provide automatic assistance in detecting software plagiarism. 

Project Martial provides a modular framework that allows easily new detection tools to be added or fine-tune the meta-parameters of the existing models to better fit. At the moment, the stack consists of three unique analysers, with more analysers expected by the end of 2024 and early 2025. 

The main analyser fits into the \textbf{SC} category and uses natural language processing techniques for computing similarities during the static analysis on the human readable parts of of the comments~\cite{math12071073}. The methodology is based on using transformers (RoBERTa~\cite{liu2019roberta} and Universal Sentence Encoder~\cite{cer2018universal}) to embed the content of comments, followed by performing detection of potential similarities, based on computing the cosine similarity. As future work in this analyzer, we want to eventually expand these techniques work on code with no human annotations, such as comments or other directives. In this space, novel code analysing techniques based on large language models have proved a good potential to help into expanding machine-native source codes with annotations in human language, essentially acting as a pre-processing phase of the source code before applying the existing model. 

The second analyser is based solely on inspection of machine-readable comments(e.g. linter directives), that computes distances in multi-dimensional spaces to compute the estimated code similarity, based on one-hot encoded representations of the source code.~\cite{math12071073}. This technique also fits into the \textbf{SC} category.

The third approach in project Martial uses dynamic-code complexities~\cite{folea2023complexity} for detecting software plagiarism and thus is a contributions to the techniques in \textbf{BIN} category. We plan on extending the existing birthmarks (currently based on a limit subset of metrics, such as CPU cycles, branch prediction or cache misses statistics) to capture more and more performance metrics as well as working on optimizing the similarity detection function.

Detecting similarity comparisons based on networking telemetry is an additional area of work. The initial approach focuses on finding commonalities between relational databases and evaluates whether internal database information can be found and compared by examining fingerprints that are taken from client-side network packets. Through the process of recording and examining the distinct features that are contained in these packets—like protocol headers, SQL query patterns, and signatures—it would be feasible to identify linkages between databases relying solely on payload content and the manner of communication.

%
%
%
\bibliographystyle{splncs04}
\bibliography{references}

\section*{Appendix A: Code listings for obfuscation techniques}

\renewcommand{\thefigure}{A\arabic{figure}}
\setcounter{figure}{0}
\renewcommand{\figurename}{Code\ Listing}

\begin{figure}[h]
\noindent\begin{minipage}[t]{.45\textwidth}
\begin{lstlisting}[language=diff]
s := []int{1, 2, 3, 4, 5, 6}
- for i := 0; i < 6; i += 1 {
    sum += s[i]
- }
\end{lstlisting}
\end{minipage}\hfill
\begin{minipage}[t]{.45\textwidth}
\begin{lstlisting}[language=diff]
s := []int{1, 2, 3, 4, 5, 6}
+ for i := 0; i < 6; i += 3 {
    sum += s[i]
+   sum += s[i+1]
+   sum += s[i+2]
+ }
\end{lstlisting}
\end{minipage}
\caption{Example of a loop unrolling obfuscation.}
\label{loop-unrolling}
\end{figure}

\begin{figure}[h]
\noindent\begin{minipage}[t]{.45\textwidth}
\begin{lstlisting}[language=diff]
  s, err := f()
- if err != nil {
-   return nil, err
- }
- // main logic
\end{lstlisting}
\end{minipage}\hfill
\begin{minipage}[t]{.45\textwidth}
\begin{lstlisting}[language=diff]
  s, err := f()
+ if err == nil {
+   // main logic
+ }
+ return nil, err
\end{lstlisting}
\end{minipage}
\caption{Example of branch inversion obfuscation.}
\label{control-flow-conditional-branches}
\end{figure}

\begin{figure}[h]
\noindent\begin{minipage}[t]{.45\textwidth}
\begin{lstlisting}[language=diff]
- if '0' <= c && c <= '9' {
-   c = byte(c - 'A' + 10)
- } elif case 'A' <= c && c <= 'F' {
-   c = byte(c - '0')
- }
- u := byte('C')
\end{lstlisting}
\end{minipage}\hfill
\begin{minipage}[t]{.45\textwidth}
\begin{lstlisting}[language=diff]
+ func unhex(c byte) byte {
+   if '0' <= c 
+     && c <= '9' {
+     return c - '0'
+   if 'A' <= c && c <= 'F' {
+      return c - 'A' + 10
+   }
+   return 0
+ } 
+ u := unhex(byte('C'))
\end{lstlisting}
\end{minipage}
\caption{Example of obfuscation of the original code by aggregating multiple instruction in a dedicated method.}
\label{aggregating-multiple-instruction}
\end{figure}

\end{document}